\documentclass[aps,prb,twocolumn,superscriptaddress,preprintnumbers]{revtex4}

\usepackage[colorlinks,bookmarks=false,citecolor=blue,linkcolor=red,urlcolor=blue]{hyperref}
\usepackage{epsfig}
\usepackage{tikz}
\usepackage{graphicx}
\usepackage{physics}
\usepackage{dcolumn}
\usepackage{bm}

\usepackage{amsmath,amssymb}

\makeatletter
\renewcommand*\env@matrix[1][*\c@MaxMatrixCols c]{%
  \hskip -\arraycolsep
  \let\@ifnextchar\new@ifnextchar
  \array{#1}}
\makeatother

\usepackage{appendix}

\begin{document}

\title{Excitonic collective modes in Weyl semi-metals}

\author{N. S. Srivatsa}
\email{srivatsa@pks.mpg.de}
\affiliation{The Institute of Mathematical Sciences, HBNI, C I T Campus, Chennai 600 113, India}
\affiliation{Max Planck Institute for the Physics of Complex Systems, N\"othnitzer Str. 38, 01187 Dresden, Germany}
\author{R. Ganesh}
\email{ganesh@imsc.res.in}
\affiliation{The Institute of Mathematical Sciences, HBNI, C I T Campus, Chennai 600 113, India}

\date{\today}

\begin{abstract}
Weyl semi-metals are three dimensional generalizations of graphene with point-like Fermi surfaces. Their linear electronic dispersion leads to a window in the particle-hole excitation spectrum which allows for undamped propagation of collective excitations. We argue that interactions in Weyl semi-metals generically lead to well-defined exciton modes. However, using a minimal model for interactions, we show that the exciton binding energy is exponentially small for weak interactions. This is due to effective two-dimensional character in the space of particle-hole pairs that are available for bound state formation. This is ultimately a consequence of linear electronic dispersion in three dimensions. Nevertheless, intermediate interaction strengths can lead to sharp spin-carrying excitonic resonances. We demonstrate this in a model Weyl semi-metal with broken time-reversal symmetry and Hubbard interactions, using GRPA (generalized random phase approximation) analysis. Excitons in Weyl semi-metals have evoked interest as their condensation could lead to an axionic charge density wave order. However, we find that the leading instability corresponds to intra-valley spin density wave order which shifts the Weyl points without opening a gap. Our results suggest interesting directions for experimental studies of three dimensional Dirac systems.

\end{abstract}
\keywords{}
\maketitle

\section{Introduction}
Dirac systems such as graphene\cite{CastroNeto2009}, Weyl semimetals\cite{Rao2016,Yan2017} and Dirac semimetals\cite{Armitage2018} are of great interest due to their point-like Fermi surfaces and conical dispersions. The effects of electron-electron interactions in these systems are especially interesting\cite{Kotov2012} with studies focussing on quasiparticle character, ordering instabilities, etc. A particularly elegant feature was pointed out by Jafari and Baskaran in the context of graphene\cite{Baskaran2002,Jafari2005}. They argued that conical dispersion leads to a window-like structure in the particle-hole continuum within which excitonic modes can propagate. 
In this article, we extend this notion to three-dimensional Weyl semi-metals. 
We show that they generically host undamped spin-carrying collective excitations.

\begin{figure}
\includegraphics[width=3.3in]{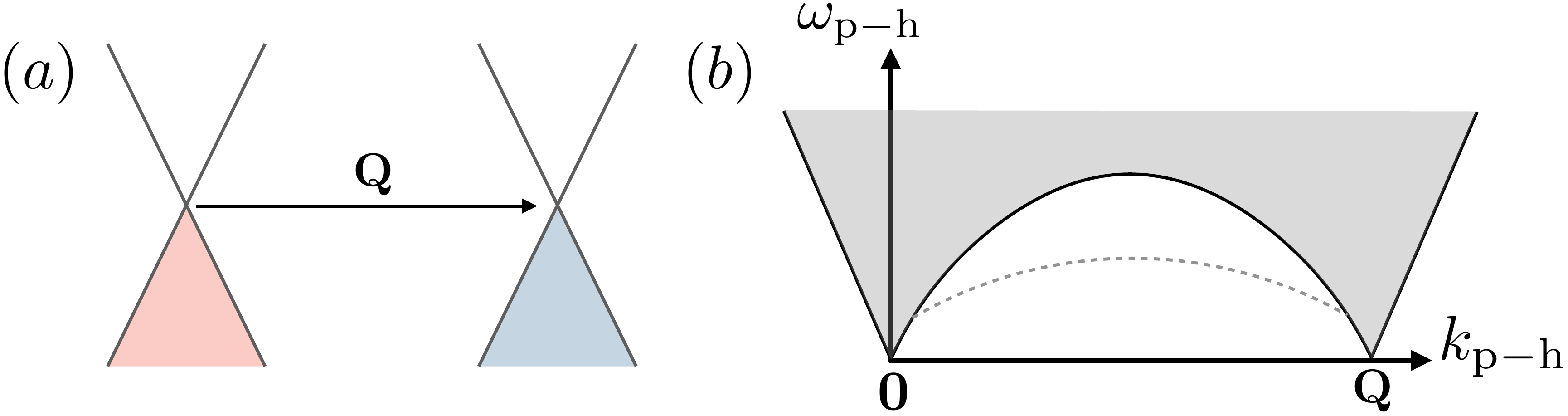} 
\caption{(a) Dispersion in a minimal Weyl semi-metal with two valleys separated by momentum $\mathbf{Q}$. (b) The corresponding particle-hole continuum. Low energy particle-hole excitations can either be intra-valley ($\mathbf{k}_{p-h} \sim \mathbf{0}$) or inter-valley ($\mathbf{k}_{p-h} \sim \mathbf{Q}$). The resulting window structure allows for the propagation of an undamped collective mode -- schematically shown as a dashed line.
}
\label{fig.schematic}
\end{figure}

The suitability of Weyl semi-metals for hosting exciton collective modes stems from their linear dispersion. This is illustrated in Fig.~\ref{fig.schematic}(a) for a simple Weyl semi-metal. It has two Weyl points which occur at incommensurate wavevectors separated by $\mathbf{Q}$. Low energy quasiparticle excitations can occur in either valley. The corresponding particle-hole continuum is shown schematically in Fig.~\ref{fig.schematic}(b). At low energies, it consists of two cones centred at momentum zero and $\mathbf{Q}$, corresponding to intra-valley and inter-valley particle-hole excitations respectively. This continuum is very different compared to that of a conventional metal, say with a spherical Fermi surface. In the latter, low energy excitations near the Fermi surface form a swathe-like particle-hole continuum, extending to zero energy over a wide range of momentum values. In contrast, the Weyl semi-metal possesses a window structure which can host collective excitations as shown in Fig.~\ref{fig.schematic}(b). Such a collective mode will remain undamped as it cannot decay into particle-hole pairs while conserving energy and momentum.

The impact of electron interactions on the stability of a Weyl semi-metal is a question of considerable interest. Due to the vanishing density of states at the Fermi level, it is readily seen that weak interactions cannot bring about instabilities. However, a sufficiently strong interaction will lead to ordering instabilities\cite{Krempa2014}. A straightforward comparison can be made with the honeycomb lattice Hubbard model, which develops an antiferromagnetic instability at a critical interaction strength\cite{Sorella2012}. 
In Weyl semi-metals, an elegant possibility is an instability to an `axionic insulator'\cite{Bitan,Wang2013,Wang2016}. This emerges as a natural inter-valley `mass' term that opens an electronic gap. 
The physics of this transition and the associated soft modes is of great interest. In particular, it has been argued that low energy behaviour in the vicinity of this transition exhibits emergent supersymmetry\cite{Jian2015}. This prompts the following question: is there a microscopic model with a tunable parameter that can realize axion condensation? We study a Hubbard model which is the simplest plausible microscopic paradigm. However, we find that the Hubbard interaction merely shifts the Weyl points and does not open a gap.

The remainder of this article is structured as follows. In Sec.~\ref{sec.GRPA}, we outline the generalized random phase approximation (GRPA) formalism that we use to find collective excitations. In Sec.~\ref{sec.simpmodel}, we consider a simplistic model for interactions in a Weyl semi-metal which allows for an analytic calculation of the collective mode spectrum. We show that linear electronic dispersion leads to an effectively two-dimensional phase space for exciton formation. 
Next, in Sec.~\ref{sec.Hubbard}, we consider Hubbard interactions in a model Weyl semi-metal. We show that excitonic modes occur with intra-valley as well as inter-valley character. Sec.~\ref{sec.Hubbardorderings} discusses exciton properties such as binding energy and spinful character. It demonstrates that exciton condensation leads to magnetic order. We conclude with a summary and discussion in Sec.~\ref{sec.summary}.

\section{Generalized random phase approximation}
\label{sec.GRPA}

\begin{figure*}
\includegraphics[width=6.6in]{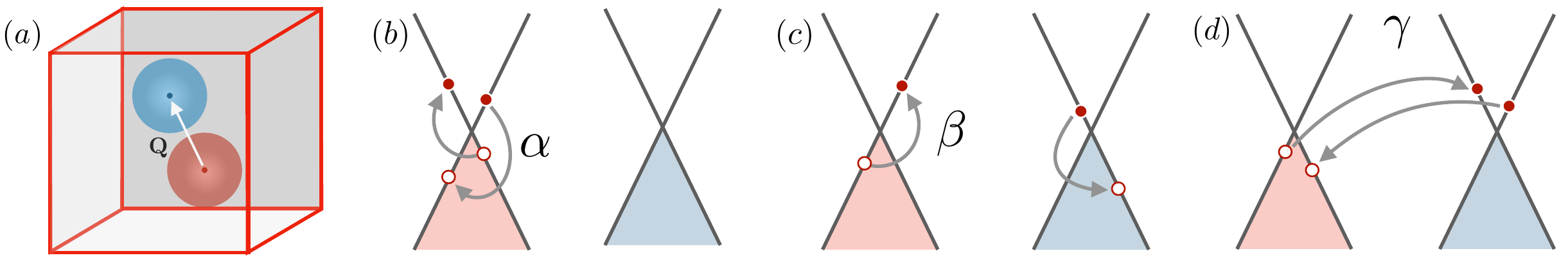}
\caption{(a) Brillouin zone for a three dimensional lattice, assumed to be cubic for simplicity. The spherical regions around Weyl points show linear dispersion. We approximate the Brillouin zone as consisting only of these spherical regions. (b-d) Interaction processes at low energies. We have processes with small momentum transfer that involve the same valley ($\alpha$) or different valleys ($\beta$), apart from large-momentum-transfer inter-valley scattering ($\gamma$). 
}
\label{fig.simpmodel}
\end{figure*}

We first describe the GRPA scheme\cite{Ganeshthesis} for finding collective excitations in general terms. In the following sections, we apply this formalism to models with increasing level of detail. This is a weak coupling approach where we begin with a non-interacting Hamiltonian, denoted by $\mathcal{H}_{KE}$. We calculate its susceptibility to various orders. We then include interaction terms which modify this `bare' susceptibility. Finally, we identify divergences in the renormalized GRPA susceptibility as collective mode resonances. This approach is equivalent to summation over ladder diagrams (e.g., compare Refs.~\onlinecite{Yunomae2009,Ganesh2009}), the Bethe-Salpeter equation\cite{Liegener1981,Koinov2010} and equations-of-motion approaches\cite{Rowe1968}. It has been shown to work well even in the strong coupling limit where it provides good agreement with the appropriate spin wave expansion\cite{Belkhir1994,Ganesh2009}. 

Starting with an appropriate non-interacting Hamiltonian, $\mathcal{H}_{KE}$, we consider various ordering tendencies represented by fermionic bilinears, e.g., $\hat{\mathbf{S}}_\mathbf{q} \equiv \sum_\mathbf{k} c_{\mathbf{k}+\mathbf{q},\mu}^\dagger \mathbf{\sigma}_{\mu,\nu}c_{\mathbf{k},\nu}$ for spin density wave order at momentum $\mathbf{q}$. 
We collect all such relevant bilinears into an array, $\hat{O}(\mathbf{q})$. Assuming fictitious fields that couple to these bilinears, we have
\begin{eqnarray}
&\mathcal{H}= \mathcal{H}_{KE} -\frac{1}{N}\sum_{\mathbf{q}}h_{\beta}(q,t)\hat{O}_{\beta}^{\dagger}(\mathbf{q}),
\label{eq.Hperta}
\end{eqnarray}
where $N$ is the number of sites in the system. 
Within linear response, these induce expectation values given by 
\begin{equation}
\expval{\hat{O}_{\alpha}}(\mathbf{q},t)=\int_{-\infty}^{\infty}dt'\chi_{\alpha\beta}^{0}(\mathbf{q},t-t')h_{\beta}(\mathbf{q},t').
\end{equation}
The `bare' susceptibility matrix, $\chi^{0}$, is computed using the spectrum of $\mathcal{H}_{KE}$ (assuming zero temperature),
\begin{equation}
\chi^{0}_{\alpha,\beta}(\mathbf{q},t-t')=i\frac{\theta(t-t')}{N}\expval{[\hat{O}_{\alpha}(\mathbf{q},t),\hat{O}_{\beta}^{\dagger}(\mathbf{q},t')]}_{0}.
\end{equation}
This can be directly evaluated in frequency space to give $\chi^{0}_{\alpha,\beta}(\mathbf{q},\omega)$, which takes a form similar to the Lindhard function. We have $\expval{\hat{O}_{\alpha}}(\mathbf{q},\omega) = \chi^{0}_{\alpha,\beta}(\mathbf{q},\omega) h_\beta(\mathbf{q},\omega)$.

We next consider interactions represented by two particle processes, denoted by $\mathcal{H}_{int}$. In the GRPA scheme, this is quadratically decoupled so as to renormalize the effective fields in Eq.~\ref{eq.Hperta}. We have
\begin{equation}
\mathcal{H}_{int} \longrightarrow g ~{\hat{O}_{\beta}^{\dagger}}.D_{\beta\alpha}.\expval{\hat{O}_\alpha},
\end{equation}
where $g$ is the interaction strength and $D_{\beta\alpha}$ is a coupling matrix. The expectation values, $\expval{\hat{O}_{\alpha}}$, can depend on space and time. These quadratically decomposed terms renormalize the coupling fields in Eq.~\ref{eq.Hperta}, leading to \begin{equation}
\expval{\hat{O}_{\alpha}}=\chi^{0}_{\alpha\beta}\left(h_{\beta} -g D_{\beta\tau}\expval{\hat{O}_{\tau}}\right).
\end{equation}
Upon rearranging the above equation, we arrive at an expression for the expectation value of the induced order,
\begin{equation}
\expval{\hat{O}_{\alpha}}(\mathbf{q},\omega)=\chi^{GRPA}_{\alpha\beta}(\mathbf{q},\omega)h_{\beta}(\mathbf{q},\omega),
\end{equation}
where 
\begin{eqnarray}
\chi^{GRPA}_{\alpha\beta}=[(1+g\chi^{0}D)^{-1}\chi^{0}]_{\alpha\beta}.
\label{eq.chiGRPA}
\end{eqnarray}
This gives the GRPA susceptibility matrix at momentum $\mathbf{q}$ and frequency $\omega$. We have suppressed $(\mathbf{q},\omega)$ arguments of $\chi^0$ and $\chi^{GRPA}$ for simplicity. Operationally, we first evaluate $\chi^0$, numerically if necessary. We then seek $(\mathbf{q},\omega)$ where $(1+g\chi^{0}D)$ becomes singular. This indicates that ordering will develop for an infinitesimal inducing field, $h_{\beta}(\mathbf{q},\omega)$. The locus of such $(\mathbf{q},\omega)$ points provides the dispersion of collective excitations.

\section{A generic interacting Weyl semimetal}
\label{sec.simpmodel}

To develop a simple model for a Weyl semi-metal, we consider the system shown in Fig.~\ref{fig.schematic}(a). At low energies, this system is described by a particularly simple single-particle Hamiltonian,
\begin{eqnarray}
 \mathcal{H}_{KE}^{simp} =   \sum_{\bf k} \Psi_{\mathbf{k} }^\dagger H_\mathbf{k} \Psi_{\mathbf{k}},
\label{eq.HKE}
\end{eqnarray}
where $\Psi_{\mathbf{k}}^\dagger = \left(
\begin{array} {cccc}
c_{c,L,\mathbf{k}}^\dagger & c_{c,R,\mathbf{k}}^\dagger & c_{v,L,\mathbf{k}}^\dagger & c_{v,R,\mathbf{k}}^\dagger   
\end{array}
\right)$ is the array of quasiparticle creation operators. They are defined in the band basis with the index $c/v$ denoting conduction/valence bands. The $L/R$ indices represent the two valleys, with $\mathbf{k}$ representing deviation from the corresponding Weyl point. In this basis, the Hamiltonian matrix takes a simple form, $H_\mathbf{k}= \mathrm{Diag}\{ +\vert \mathbf{k}\vert  ,+\vert \mathbf{k} \vert ,-\vert \mathbf{k} \vert ,-\vert \mathbf{k}\vert   \}$. We set the Fermi velocity to unity and the chemical potential to zero, as appropriate for an isotropic undoped Weyl semi-metal. 

With the goal of developing a minimal model, we assume a simple form for interactions as shown in Fig.~\ref{fig.simpmodel}(b-d). At low energies, two-particle processes can only be of two types -- with low (comparable to zero) or high (comparable to $\mathbf{Q}$)  momentum transfer. Low-momentum-transfer processes can be further subdivided into two classes -- within a single valley or those involving both valleys. As a simplifying assumption, we take these processes to have momentum-independent amplitudes, given by $\alpha$, $\beta$ and $\gamma$ as shown in the figure. This leads to the Hamiltonian,
\begin{eqnarray}
\nonumber \mathcal{H}_{int}^{simp.} &=& -\frac{\alpha}{N} \sum_{\mathbf{k},\mathbf{k'},\mathbf{q}} \sum_{\mu = R,L} c_{c,\mathbf{k}+\mathbf{q},\mu}^\dagger c_{v,\mathbf{k}'-\mathbf{q},\mu}^\dagger c_{c,\mathbf{k}',\mu} c_{v,\mathbf{k},\mu} \\
\nonumber  &-& \frac{\beta}{N} \sum_{\mathbf{k},\mathbf{k'},\mathbf{q}}  c_{c,\mathbf{k}+\mathbf{q},R}^\dagger c_{v,\mathbf{k}'-\mathbf{q},L}^\dagger c_{c,\mathbf{k}',L} c_{v,\mathbf{k},R} \\
\nonumber &-&
\frac{\beta}{N} \sum_{\mathbf{k},\mathbf{k'},\mathbf{q}}  c_{c,\mathbf{k}+\mathbf{q},L}^\dagger c_{v,\mathbf{k}'-\mathbf{q},R}^\dagger c_{c,\mathbf{k}',R} c_{v,\mathbf{k},L} \\
 \nonumber &-& \frac{\gamma}{N} \sum_{\mathbf{k},\mathbf{k'},\mathbf{q}}  c_{c,\mathbf{k}+\mathbf{q},R}^\dagger c_{v,\mathbf{k}'-\mathbf{q},L}^\dagger 
c_{c,\mathbf{k}',R} 
c_{v,\mathbf{k},L} \\
&-& \frac{\gamma}{N} \sum_{\mathbf{k},\mathbf{k'},\mathbf{q}}  c_{c,\mathbf{k}+\mathbf{q},L}^\dagger c_{v,\mathbf{k}'-\mathbf{q},R}^\dagger 
c_{c,\mathbf{k}',L} 
c_{v,\mathbf{k},R} .
\label{eq.Hint}
\end{eqnarray}
This form of the interaction Hamiltonian is admittedly simplistic. Nevertheless, it allows for an analytic calculation of the collective mode spectrum, which in turn brings out essential aspects of the problem.

\subsection{Evaluating bare susceptibility}

We now apply the GRPA formalism taking the non-interacting Hamiltonian to be that in Eq.~\ref{eq.HKE} and the interactions to be given by Eq.~\ref{eq.Hint}.
The ordering tendencies in this system are represented by excitonic bilinears of the form $\hat{\rho}_{\eta\lambda} (\mathbf{q}) = \sum_{\mathbf{k}} 
c_{c,\eta,\mathbf{k} + \mathbf{q}}^\dagger c_{v,\lambda,\mathbf{k}} $, where $\eta,\lambda = L/R$ are valley-indices. We organize these $\hat{\rho}_{\eta\lambda}$ operators into a vector, as in Eq.~\ref{eq.Hperta}, to give
\begin{eqnarray}
 \hat{O}(\mathbf{q})=\left(\begin{array}{cccc} \hat{\rho}_{LL}(\mathbf{q}) & \hat{\rho}_{RR}(\mathbf{q}) & \hat{\rho}_{LR}(\mathbf{q}) & \hat{\rho}_{RL}(\mathbf{q}) \end{array}\right). 
 \label{eq.Oarray}
 \end{eqnarray}

The bare susceptibility to these orders takes the Lindhard form,
\begin{eqnarray}
\chi_{\alpha\beta,\eta\lambda}^{0}(\mathbf{q}, \tilde{\omega})=\sum_{\mathbf{k}}\frac{\delta_{\alpha\eta}\delta_{\beta\lambda}}{(E_{\mathbf{k}+\mathbf{q}}+E_{\mathbf{k}}-\tilde{\omega}+i0^+)},
\label{eq.chibare}
\end{eqnarray}
where $E_{\mathbf{p}} = \vert\mathbf{p}\vert$ is the quasiparticle energy and $\tilde{\omega}\equiv-\omega$. We have added an infinitesimal in the denominator for regularization. Since the bare susceptibility is diagonal in the $\hat{\rho}$ basis, we drop the indices and simply refer to it as $\chi^{0}(\mathbf{q},\tilde{\omega})$. We evaluate this quantity assuming that (a) the Brillouin zone can be approximated as consisting of two spheres of radius $k_c$, each centred at a Weyl point, and (b) the linear dispersion around each Weyl point, as given in Eq.~\ref{eq.HKE}, extends over each entire sphere. These approximations can be justified by noting that the dominant contribution to the sum in Eq.~\ref{eq.chibare} comes from the immediate neighbourhood of each Weyl point -- our approach indeed retains the correct quasiparticle energies here. In this picture, the particle-hole continuum has an intra-valley and an inter-valley component, both of which are bounded from below by the cone $\tilde{\omega} = \vert \mathbf{q}\vert$. The susceptibility is the same for both intra- and inter-valley sectors.

\begin{figure}
\includegraphics[width=2.7in]{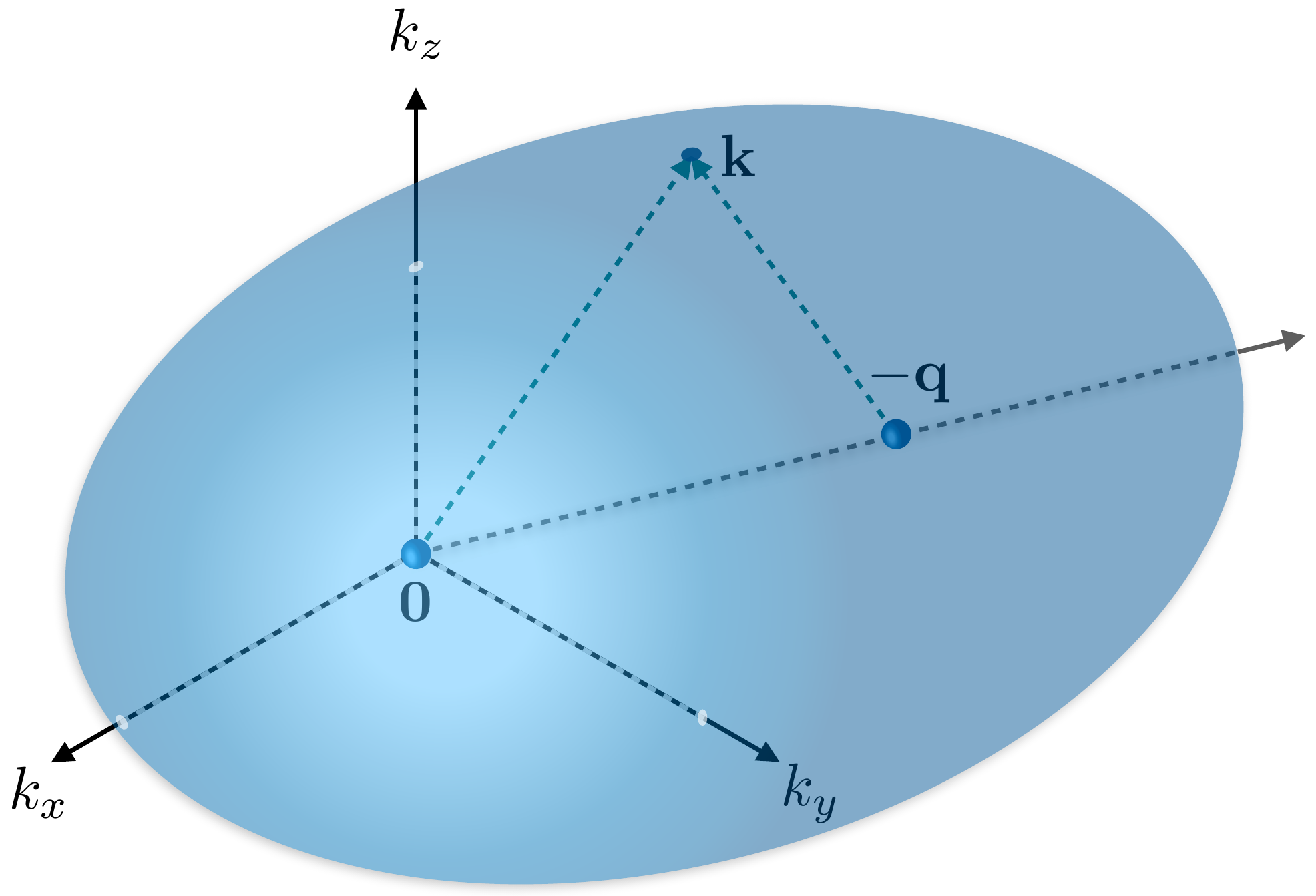}
\caption{Locus of points that contribute to $\operatorname{Im}(\chi^{0}(\mathbf{q},\tilde{\omega}))$, defined by $\vert \mathbf{k} \vert + \vert \mathbf{k}+\mathbf{q} \vert = \tilde{\omega} $. Any point on the surface has the sum of distances from the $\mathbf{0}$ and $-\mathbf{q}$ equal to $\tilde{\omega}$. These points form an ellipsoid in momentum space with foci at $\mathbf{0}$ and $-\mathbf{q}$. The major axis is along $\mathbf{q}$, with the semi-major axis being $\tilde{\omega}/2$. The eccentricity is given by $\vert \mathbf{q} \vert/\tilde{\omega}$.}
\label{fig.ellipsoid}
\end{figure}

The susceptibility in Eq.~\ref{eq.chibare} can be evaluated using an elegant geometric picture, by noting that $\operatorname{Im}(\chi^{0}(\mathbf{q},\tilde{\omega}))$ only receives contributions from $\mathbf{k}$-points which satisfy $E_{\mathbf{k}+\mathbf{q}}+ E_\mathbf{k} = \tilde{\omega}$, i.e., $\vert\mathbf{k}+\mathbf{q}\vert+ \vert\mathbf{k}\vert = \tilde{\omega}$. This relation describes an ellipsoid in $\mathbf{k}$-space with major axis along $\mathbf{q}$, as shown in Fig.~\ref{fig.ellipsoid}. In fact, $\operatorname{Im}(\chi^{0}(\mathbf{q},\tilde{\omega}))$ simply counts the number of $\mathbf{k}$ points that lie on this surface. Details of the derivation are presented in Appendix \ref{app.bare}. We find
 \begin{eqnarray}
 \operatorname{Im}(\chi^{0}(\mathbf{q},\tilde{\omega}))\approx  \left\{ \begin{array}{c}
 \frac{\pi(3\tilde{\omega}^2 - q^2)}{6}, q < \tilde{\omega} < (2k_c+q)\\
 0, \phantom{abc}\mathrm{otherwise}
 \end{array}\right.,
 \end{eqnarray}
 which vanishes outside the particle-hole continuum. This quantity, near the bottom of the particle-hole continuum ($\tilde{\omega} \gtrsim q$), can be understood as the density of `free' particle-hole pairs that are available for exciton formation. 
Remarkably, this expression reveals a quasi-two-dimensional character in the problem. As we approach the bottom of the particle-hole continuum from above ($\tilde{\omega}\rightarrow q^+$), we find that $ \operatorname{Im}(\chi^{0}(\mathbf{q},\tilde{\omega})) \rightarrow  \pi q^2/3 $, a constant for a given $\mathbf{q}$. This is analogous to the density of states of a conventional two-dimensional metal. 
This indicates that exciton formation here is analogous to bound state formation in a conventional two-dimensional system, even though we are concerned with a three dimensional Weyl semi metal.

 Using the Kramers-Kronig relation (see Appendix \ref{app.bare}), we obtain
 \begin{eqnarray}
 \operatorname{Re}(\chi^{0}(\mathbf{q},\tilde{\omega}))\approx   \pi k_c^2+\frac{\pi}{6}\left[12 q k_{c}-2q^2\log\left\{\frac{q-\tilde{\omega}}{2k_c}\right\}\right].\phantom{ab}
 \label{eq.rechi}
 \end{eqnarray}
This expression holds near the upper boundary of the window in particle-hole continuum ($\tilde{\omega} \lesssim q$). This is the region in the $\tilde{\omega}-\mathbf{q}$ space that can develop collective excitations in the presence of weak interactions.
  
 \subsection{Collective mode spectrum} 
Having found the bare susceptibility, we take interactions into account. The terms in Eq.~\ref{eq.Hint} can be quadratically decomposed as
\begin{eqnarray}
\mathcal{H}_{int}^{simp} \longrightarrow \sum_{\mathbf{q}}\expval{\hat{O}(\mathbf{q})}.D^{simp}.{\hat{O}^{\dagger}(\mathbf{q})},
\label{eq.Hdec}
\end{eqnarray}
where $\hat{O}$ has been defined in Eq.~\ref{eq.Oarray} and
\begin{eqnarray}
D^{simp} = (-)\begin{bmatrix}
\alpha & \beta&0&0\\
\beta & \alpha&0&0\\
0&0&\gamma&0\\
0&0&0&\gamma
\end{bmatrix}.
\end{eqnarray}
The coefficients $\alpha$, $\beta$ and $\gamma$ are the interaction amplitudes described in Fig.~\ref{fig.simpmodel}(b-d).
Following the GRPA prescription, we obtain
\begin{eqnarray}
\nonumber \chi^{GRPA}_{\alpha\beta,\eta\lambda}(\mathbf{q},\tilde{\omega}) \!\!= \!\!\left\{\!\left[  1+ D^{simp} \chi^{0}(\mathbf{q},\tilde{\omega})\right]^{-1}\!\right\}_{\!\alpha\beta,\mu\zeta}\!\!  \chi^{0}_{\mu\zeta,\eta\lambda}(\mathbf{q},\tilde{\omega}).\phantom{A}
\end{eqnarray}
A collective mode emerges when $\chi^{GRPA}(\mathbf{q},\tilde{\omega}) $ diverges. This occurs when $ \det[ 1+ D^{simp} \chi^{0}(\mathbf{q},\tilde{\omega})] =0$, leading to the following four solutions,
\begin{equation}
\tilde{\omega}_{i}=q-2 k_c\exp{\frac{6(qk_c-t_i)}{q^2}}, i={1,2,3,4},
\label{eq.cmdisp}
\end{equation}
where $t_i$ are given by
\begin{eqnarray}
t_{1,2}=\frac{\frac{1}{\alpha \pm \beta}-\pi k_c^2}{2\pi}, \phantom{ab}
t_{3,4}=\frac{\frac{1}{\gamma}-\pi k_c^2}{2\pi}.
 \label{eq.ts}
\end{eqnarray}
These expressions come with the following caveat. They are derived from Eq.~\ref{eq.rechi} which is only valid immediately below the particle-hole continuum, i.e., for $\tilde{\omega} \rightarrow q^-$. If they lead to a solution in this region of $\tilde{\omega}-q$ space, it represents a true collective mode. Solutions outside this region are spurious and not physically meaningful.

\subsection{Exciton binding energy}
In the right hand side of Eq.~\ref{eq.cmdisp}, the second term encodes binding energy of the collective mode, i.e., the separation from the bottom of the continuum. To have meaningful collective excitations as $\mathbf{q} \rightarrow 0$, $t_i$'s must be positive. Otherwise, the binding energy grows without bound as $\mathbf{q}$ approaches zero. For example, if $\gamma>0$ and $\gamma^{-1} \gg \pi k_c^2$, we see that $t_{3,4}$ is positive. This indicates that $\tilde{\omega}_3$ is a true collective mode as $\mathbf{q}\rightarrow 0$ with an exponentially small binding energy. If we now tune $\gamma$ to stronger values, the binding energy will increase and the collective mode will shift downwards. When $\gamma$ reaches a critical value, $\gamma_c= (\pi k_c^2)^{-1}$, $t_{3,4}$ vanishes. This indicates an instability of the Weyl semi-metal to inter-valley exciton condensation (as $\gamma$ is an intervalley process). Beyond this point, the binding energy grows sharply, indicating softening of the collective mode. More generally, the $t_i$'s in Eq.~\ref{eq.ts} encode critical interaction strengths at which instabilities arise.

We have argued above that exciton formation here is a problem of bound state formation in effectively two dimensions. In this respect, it is directly analogous to the well-known Cooper pair\cite{Cooper1956} problem, e.g., as described in Ref.~\onlinecite{Esebbag1992}. The two-dimensionality arises as only a thin shell around the Fermi surface  considered. This leads to a constant density of states, $g(\epsilon_F)$. For weak interactions encoded by $V$, a bound state is formed with an exponentially small binding energy given by $E_{Cooper} = 2\hbar \omega_D \exp(-2/g(\epsilon_F) V)$. This expression closely matches our result in Eq.~\ref{eq.cmdisp}. For concreteness, let us consider $i=3$ with $\gamma>0$ in the limit $q\rightarrow 0$. For weak interactions�($\gamma^{-1} \gg \pi k_c^2$), the binding energy comes out to be $E_b \approx 2k_c \exp(-3/\pi q^2 \gamma)$. This has precisely the same form as $E_{Cooper}$; the $q^2$ in the exponent arises from the density of states of free particle-hole states (see $\operatorname{Im}(\chi^{0}(\mathbf{q},\tilde{\omega}))$  above). 

We have demonstrated that excitons in Weyl semi-metals are analogous to Cooper pairs in metals. The existence of bound Cooper pair solutions indicates an instability of the Fermi surface, showing that metals are generically unstable to superconductivity. Likewise, the particle-hole continuum in Weyl semi-metals is unstable to exciton formation. This shows that Weyl semi-metals will generically host excitonic modes.

\section{Weyl semi-metal with the Hubbard interaction}
\label{sec.Hubbard}
In the previous section, we have considered a minimal model of an interacting Weyl semi-metal and derived analytic expressions for the collective mode spectrum. Here, we take a somewhat more realistic approach with a microscopically motivated Weyl Hamiltonian and on-site interactions. Following Burkov and Balents\cite{Burkov2011} (BB), we work with the non-interacting Hamiltonian,  
\begin{equation}
\mathcal{H}_{KE}^{BB}= v_{f}(\hat{z}\times {\pmb\sigma})\cdot \mathbf{k}+ m(k_{z})\sigma^{z},
\label{eq.H_TINI}
\end{equation}
where $(\sigma^{x}, \sigma^{y}, \sigma^{z})$ are the usual Pauli matrices and $m(k_{z})=b-t(k_{z})$, with $t(k_{z})=\sqrt{t_{S}^2+t_{D}^2+2t_{S}t_{D}\cos({k_{z}d})}$.
This model was derived by considering a topological insulator - normal insulator superlattice with broken time reversal symmetry. The quantities $( t_{	S} , t_{D} $) represent effective hopping amplitudes in the heterostructure, while $b$ is the time-reversal breaking term. It realizes a 2-band model in which the two-components of the wavefunction are the physical spin of the electron. The dispersion hosts Weyl points at momenta ${(0,0,P_{1,2})}$, where
\begin{equation}
P_{1,2}=\pm\frac{\pi}{d} \mp \frac{1}{d} \cos^{-1}\left({\frac{t_{S}^2+t_{D}^2-b^2}{2t_{S}t_{D}}}\right).
\end{equation}
The length scale $d$ denotes the separation between layers. For later convenience, we define $\mathbf{Q} = {(0,0,P_{1}-P_2)}$, the vector separation between the two Weyl points.
The low energy excitations here are similar to the schematic in Fig.~\ref{fig.schematic} (a,b). In particular, the particle-hole continuum has two distinct low energy regions -- intra-valley (momentum near zero) and inter-valley (momentum near $\mathbf{Q}$). 

We take the interaction Hamiltonian to be of the Hubbard form, given by
\begin{eqnarray}
\mathcal{H}_{int}^{Hubbard} = \frac{U}{N} \sum_{\mathbf{k},\mathbf{k}',\mathbf{p}} c_{\mathbf{k}+\mathbf{p},\uparrow}^\dagger c_{\mathbf{k}'-\mathbf{p},\downarrow}^\dagger c_{\mathbf{k}',\downarrow} c_{\mathbf{k},\uparrow}.
\label{eq.Hubbardint}
\end{eqnarray}

The GRPA analysis of this problem takes different forms for intra-valley ($\mathbf{q}\sim 0$) and inter-valley sectors ($q \sim \mathbf{Q}$). In particular, the decomposition of the interaction is different in the two cases. We discuss these separately below.

\subsection{GRPA in the inter-valley sector}
\label{sec.inter}
Focussing on large momenta, low energy excitations involve a particle from one valley and a hole from the other. 
 To handle this structure, we divide the Brillouin zone into two regions. We label the $k_z > 0$ region as `right' (R) and $k_z < 0$ as `left' (L). The `right' and `left' valleys contain the Weyl points $(0,0,P_1)$ and $(0,0,P_2)$ respectively.
 We define creation/annihilation operators at low energies accordingly, e.g., $c_{L,\mathbf{p},\sigma}^\dagger$ denotes creation at momentum $\mathbf{p}$ lying below the $k_x-k_y$ plane. We consider bilinears of the form
\begin{eqnarray}
\nonumber  \hat{\rho}_{inter}(\mathbf{q})&=&\frac{1}{2}\sum_{\mathbf{k}}\left\{c_{R,\mathbf{k}+\mathbf{q},\uparrow}^{\dagger}c_{L,\mathbf{k},\uparrow}+c_{R,\mathbf{k}+\mathbf{q},\downarrow}^{\dagger}c_{L,\mathbf{k},\downarrow}\right\},\\
\hat{S}_{inter}^{u}(\mathbf{q})&=&\frac{1}{2}\sum_{\mathbf{k},\mu,\mu'}c_{R,\mathbf{k}+\mathbf{q},\mu}^{\dagger}  
\sigma^u_{\mu,\mu'} c_{L,\mathbf{k},\mu'}.
\end{eqnarray}
As we are interested in inter-valley excitations, the net momentum $\mathbf{q}$ is restricted to values near $\mathbf{Q}$.
The index $u=x,y,z$ denotes three possible spin directions. As in Eq.~\ref{eq.Hperta}, we gather these bilinears into an array, $\hat{O}_{inter}(\mathbf{q})=[\hat{\rho},\hat{S}_{inter}^{z},\hat{S}_{inter}^{+},\hat{S}_{inter}^{-},\hat{\rho}^{\dagger},\{\hat{S}_{inter}^{z}\}^{\dagger},\{\hat{S}_{inter}^{+}\}^{\dagger},\{\hat{S}_{inter}^{-}\}^{\dagger}]$. 

Naively, we could have only considered spin-carrying bilinears, $\hat{S}_{inter}^{u}(\mathbf{q})$, as repulsive interactions are known to favour spin-carrying collective modes. However, due to inherent spin-orbit coupling, there is no spin rotational symmetry in Eq.~\ref{eq.H_TINI}. As a consequence, at the level of bare susceptibility, $\hat{S}_{inter}^{u}(\mathbf{q})$ and $\hat{\rho}_{inter}(\mathbf{q})$ are mixed.

Decoupling the Hubbard interaction of Eq.~\ref{eq.Hubbardint} in terms of these bilineras, we obtain the coupling matrix $D_{\alpha\beta}^{inter} = Diag\{2,-2,-1,-1,2,-2,-1,-1\}$ (details in Appendix \ref{app.GRPAinter}). We use these expressions in the GRPA formalism to find collective modes.

The bare susceptibility is an $8\times 8 $ matrix, see Appendix~\ref{app.GRPAinter} for explicit expressions. The elements of this matrix can only be found numerically. We evaluate them by discretizing the cubic Brillouin zone into a $L\times L\times L$ mesh, with $L$ up to 20. The singularities that occur at the Weyl point (the denominators in $\chi^0$ vanish here) are avoided by choosing parameters such that the Weyl points do not lie on the $\mathbf{k}$-mesh. An illustrative result is shown in Fig.~\ref{fig.cmodepeaks}(a). It plots the $\hat{S}_{inter}^z-\hat{S}_{inter}^z$ component of the $\chi^{GRPA}(\mathbf{q},\omega)$ matrix vs. $\omega$. The momentum $\mathbf{q}$ is kept fixed at a point in the vicinity of $\mathbf{Q}$. We see a clear divergent response, indicating a collective mode. This is brought about by one eigenvalue of $[1+U\chi^0_{inter}D^{inter}]$ vanishing at this point. As shown in the figure, the collective mode shifts downwards as interaction $U$ is increased.

\subsection{GRPA in the intra-valley sector}
In the intra-valley sector, we define bilinear operators
\begin{eqnarray}
\nonumber  \hat{\rho}_{\nu = L/R}(\mathbf{q})&=&\frac{1}{2}\sum_{\mathbf{k}}\left\{c_{\nu,\mathbf{k}+\mathbf{q},\uparrow}^{\dagger}c_{\nu,\mathbf{k},\uparrow}+c_{\nu,\mathbf{k}+\mathbf{q},\downarrow}^{\dagger}c_{\nu,\mathbf{k},\downarrow}\right\},\\
\hat{S}_{\nu=L/R}^{u}(\mathbf{q})&=&\frac{1}{2}\sum_{\mathbf{k},\mu,\mu'}c_{\nu,\mathbf{k}+\mathbf{q},\mu}^{\dagger}  
\sigma^u_{\mu,\mu'} c_{\nu,\mathbf{k},\mu'}.
\end{eqnarray}
The momentum $\mathbf{q}$ is taken to be small, with $\vert \mathbf{q} \vert \ll \vert \mathbf{Q}\vert$. The appropriate form of the bilinear array here is  $\hat{O}_{intra}(\mathbf{q})=[\hat{\rho}_L,\hat{S}_{L}^{z},\hat{S}_{L}^{+},\hat{S}_{L}^{-},\hat{\rho}_R,\hat{S}_{R}^{z},\hat{S}_{R}^{+},\hat{S}_{R}^{-}]$. The Hubbard interaction can be decoupled in terms of this array with the coupling matrix $D_{\alpha\beta}^{intra}=Diag\{2,-2,-1,-1\}\otimes\begin{pmatrix}
1 & 1 \\
1 & 1
\end{pmatrix}$ (details in Appendix \ref{app.GRPAintra}). 
We evaluate $\chi^{GRPA}$ numerically as described in Sec.~\ref{sec.inter} above. The resulting collective mode resonances are shown in Fig.~\ref{fig.cmodepeaks}(b,c). The plots show the $\hat{S}_{L}^z-\hat{S}_{L}^z$ and $\hat{S}_{L}^+-\hat{S}_{L}^+$ components of $\chi^{GRPA}$. The divergent peak indicates a collective mode which shifts downwards with increasing $U$.

\section{Excitons from Hubbard interactions }
\label{sec.Hubbardorderings}

As discussed above, we find collective modes both in the intra-valley and inter-valley sectors. We elaborate on some aspects of the observed exciton modes below.

\subsection{Binding energy}

Relatively large interaction strengths are required to see collective modes that are well separated from the continuum. 
In the intra-valley sector, we see clear modes only for $U\gtrsim 5$ when $t_S$, $t_D$, $b$ are close to unity (bandwidth $\sim$ 4). In the inter-valley sector, we require $U\gtrsim 8.5$. For comparison, the honeycomb lattice Hubbard model shows well-separated collective modes even for $U\sim 2$ when $t$ is unity (bandwidth $\sim$ 6) \cite{Tsuchiya2013}.
This can be understood from our analysis in Sec.~\ref{sec.simpmodel}. The effective two-dimensional phase space of excitons leads to an exponentially small binding energy, thereby requiring a large interaction strength.

\subsection{Spin character}

\begin{figure}
\includegraphics[width=3.3in]{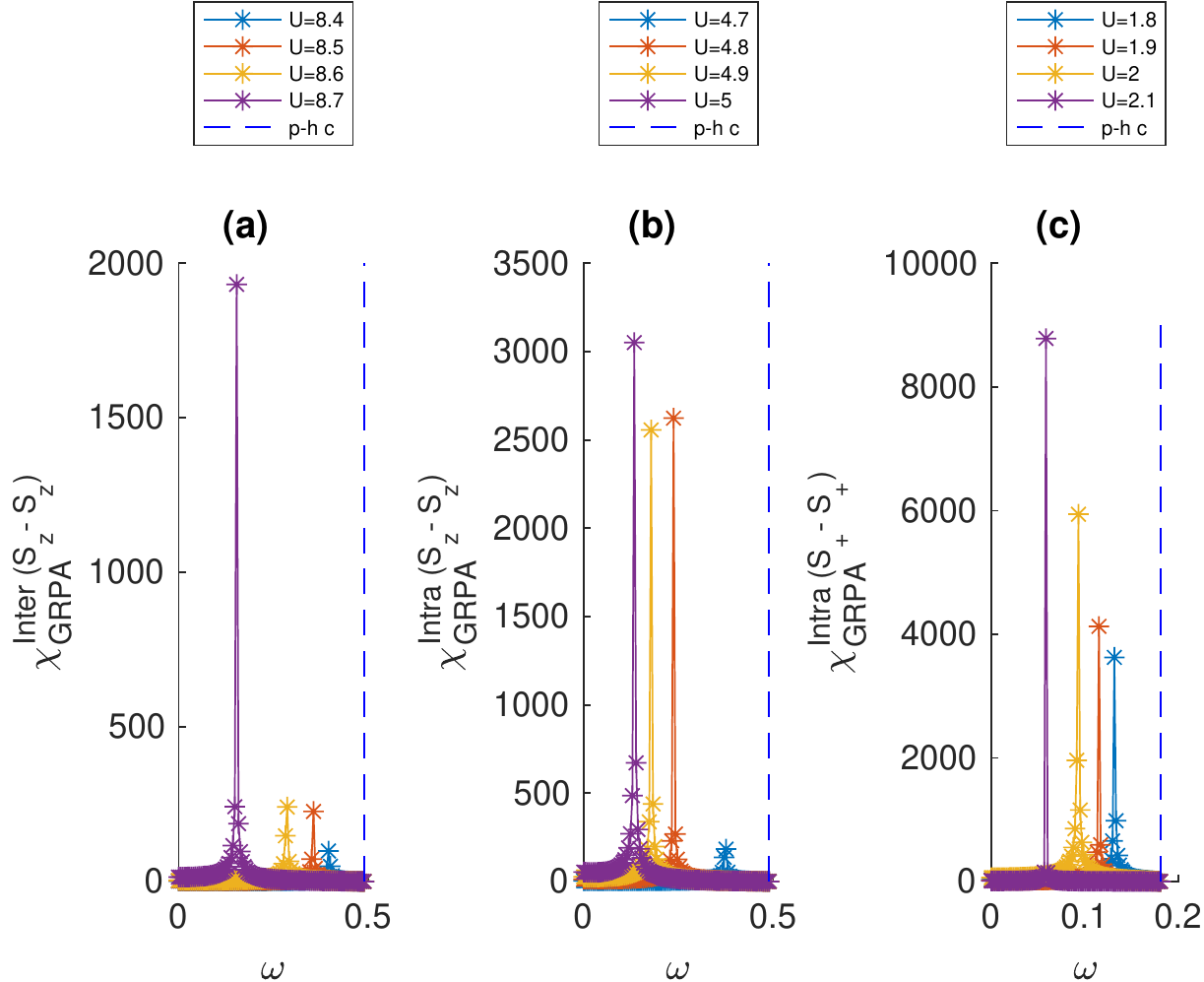}
\caption{Excitonic resonances manifested in spin-spin response calculated within GRPA. (a and b) Inter-valley and intra-valley $\hat{S}_z$-$\hat{S}_z$ response as a function of frequency. The parameters used are $t_S=1, t_D=0.9, b=1, v_f=1$. (c) Intra-valley $\hat{S}_+$-$\hat{S}_+$ response for $t_S=1, t_D=0.9, b=1, v_f=0.25$. In (a), the response is calculated at a momentum that is close to $\mathbf{Q}$. In (b) and (c), the momentum is close to zero. The dashed lines in each plot show the onset of the particle-hole continuum.}
\label{fig.cmodepeaks}
\end{figure}

In both the intra-valley and inter-valley sectors, the collective modes carry spin. At resonant ($\mathbf{q}$,$\omega$), we find large spin-spin components in the $\chi^{GRPA}_{\alpha\beta}(\mathbf{q},\omega)$ matrix as shown in Fig.~\ref{fig.cmodepeaks}. In contrast, the density-density ($\hat{\rho}$-$\hat{\rho}$) components are negligible. 
We find two distinct collective modes: one with dominant $\hat{S}^z$ character and other with $\hat{S}^\pm$ character. The latter is doubly degenerate, representing magnetic moment along $x$ and $y$ directions. 
In most regions of parameter space, only one of these modes is well separated from the continuum. Depending on $t_S$, $t_D$, $b$ and $v_F$, it is either the $\hat{S}^z$ mode or the $\hat{S}^\pm$ mode. Indeed, the anisotropy between $z$ and in-plane spin components is inherited from $\mathcal{H}_{KE}^{BB}$ in Eq.~\ref{eq.H_TINI}, which has different Fermi velocities in $z$ and in-plane directions.

 
\subsection{Exciton condensation}
An exciting prospect in an interacting Weyl semi-metal is the occurrence of an axionic insulator. In each valley, the Weyl semi-metal Hamiltonian has a Clifford algebra structure with three Pauli matrices occurring in the Hamiltonian. At the level of a single valley, no perturbation can open a gap. However, taking both valleys together, there exists a mass term that opens a full gap. The resulting state is called the axionic insulator and has several interesting properties, including defects that carry gapless excitations\cite{Wang2013}. As mass terms lead to large energy lowering by opening a full gap, one may expect that introducing interactions in a Weyl semi-metal will lead to an axionic insulator. Such a transition has been argued to possess emergent supersymmetry with the collective modes and the electronic excitations acquiring the same group velocity\cite{Jian2015}. Motivated by these arguments, we look for instabilities that arise from the Hubbard interaction. Within our GRPA approach, an instability will manifest as `softening' of a collective mode with its energy going to zero at some momentum $\mathbf{q}_{inst}$.

\begin{figure}
\includegraphics[width=3.3in]{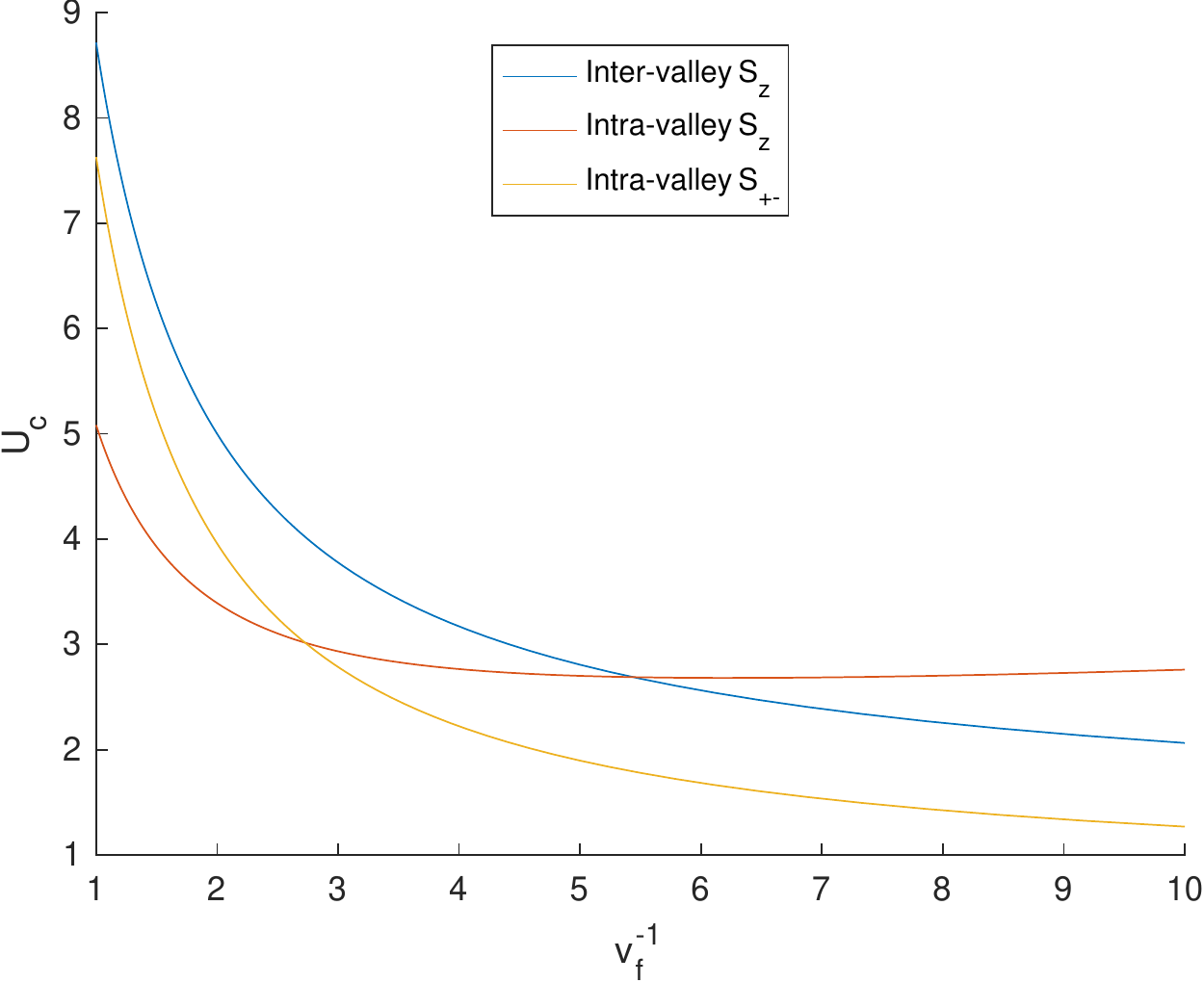}
\caption{Critical strengths ($U_c$) of intra-valley and inter-valley orderings at different values of  $v^{-1}_{F}$. We have fixed $t_S=b =1$ and $t_D=0.9$. We do not show the inter-valley $\hat{S}^\pm$ instability as it occurs at higher $U$ than the others.}
\label{fig.Critstrengths}
\end{figure}

Surprisingly, we find that the Hubbard interaction does not lead to an axionic insulator. As we increase $U$, we find that collective modes soften in the intra-valley sector, at $\mathbf{q}=0$. As the collective modes carry spin, we identify this as a magnetic instability. Depending on the parameters of in $\mathcal{H}_{KE}^{BB}$, we find two regimes (we set $t_S=b =1$ and $t_D=0.9$ for concreteness):  
(a) for $v_F^{-1}<2.8$, the leading instability is to spin ordering in the z-direction, and (b) for $v_F^{-1}>2.8$, the leading instability is to ordering in the XY plane. This is shown in Fig.~\ref{fig.Critstrengths} which shows the critical interaction strength required for exciton condensation. The figure shows critical $U$ values for three different instabilities, (i) intra-valley $\hat{S}^z$ ordering, (ii) intra-valley $\hat{S}^\pm$ ordering, and (iii) inter-valley $\hat{S}^z$ ordering. For each value of $v_F$, it is the smallest of these critical U's that has physical significance. Beyond this $U_c$, the Weyl semi-metal is unstable to magnetic order. We have independently confirmed these $U_c$ estimates by performing a mean-field calculation for each magnetic order (intra-valley $\hat{S}^z$, intra-valley $\hat{S}^\pm$ and inter-valley $\hat{S}^z$). In each case, a self-consistent magnetization emerges only when $U$ is increased beyond the corresponding critical value given by the GRPA analysis.

For any choice of parameters in $\mathcal{H}_{KE}^{BB}$, we find that the leading instability is always to intra-valley ordering. This does not open a gap in the electron dispersion. Rather, it merely shifts the Weyl points. 
For example, when $v_f=1$ in Fig.~\ref{fig.Critstrengths}, the Weyl semi-metal is stable until $U\approx 5.2$ where an excitonic mode with intra-valley $\hat{S}^z$ character softens. This indicates that an axionic insulator does not emerge from Hubbard interactions.

\section{Summary and Discussion}
\label{sec.summary}

We have discussed excitonic modes in Weyl semi-metals. Our starting point is the observation of a window in the particle-hole continuum that is conducive to the propagation of undamped collective modes. 
A similar window structure was pointed out by Baskaran and Jafari in the context of graphene. They argued that repulsive interactions in graphene naturally give rise to spin-1 (triplet) excitonic modes within this window. We have shown that these arguments extend to the three dimensional case of Weyl semi-metals. The window structure forbids the decay of collective excitations into particle-hole pairs. `Bosonic' damping is still possible via decay into pairs of collective-excitations. However, this is a more subtle effect that we do not discuss here. Our study of charge-neutral spin-carrying exciton modes serves a counterpoint to earlier work on charged plasmonic collective modes in Weyl semi-metals\cite{Ahn2016,Hofmann2016,Song2017,Sdas,Red}.

Considering a simplistic model of an interacting Weyl semi-metal, we show that low energy particle-hole excitations have an effective two-dimensional character. They lie on the surface of an ellipsoid in momentum space, constrained by linear single-particle-dispersion and energy-momentum conservation. This effective two-dimensionality leads to an exponentially small binding energy for excitons. As a consequence, a large interaction strength is required to see excitons that are well separated from the particle-hole continuum.

The approach of Baskaran and Jafari in Refs.~\onlinecite{Baskaran2002,Jafari2005} was criticised\cite{Peres2004} for not including sublattice character present in the microscopic description. In response, Baskaran and Jafari justified their approach by invoking an effective Fermi liquid picture that is not necessarily microscopic\cite{Baskaran2004}.   
Later on, Refs.~\onlinecite{Tsuchiya2012,Tsuchiya2013} presented a GRPA analysis keeping the full microscopic structure of the honeycomb lattice Hubbard model. This does show the presence of excitonic modes. However, a critical interaction strength is required to have a well defined linear mode at $\mathbf{q}\rightarrow 0$. (The analysis in Refs.~\onlinecite{Tsuchiya2012,Tsuchiya2013} is presented in the language of the attractive Hubbard model. Nonetheless, these results also apply to the repulsive Hubbard model via a particle-hole transformation).  
Beyond this critical value, the excitons condense to give rise to an antiferromagnet. In this antiferromagnetic phase, the collective modes break up into Goldstone modes and an amplitude mode\cite{Tsuchiya2013,Lu2016}.

Our simplistic model, described in Sec.\ref{sec.simpmodel}, is analogous to the initial analysis of Baskaran and Jafari. It takes the single valley Hamiltonian to be ${Diag}\{\vert \mathbf{k} \vert, -\vert \mathbf{k}\vert\}$, rather than $\mathbf{k}\cdot\mathbf{\sigma}$. As a consequence, it ignores the coherence factors that enter the eigenvectors of the single-particle Hamiltonian. Nevertheless, this analysis provides valuable insight by highlighting the effective two-dimensionality of the phase space of particle-hole pairs. Furthermore, it shows that Weyl semi-metals will generically host excitonic modes.
We go beyond this picture with a microscopic model in Sec.~\ref{sec.Hubbard}, providing a full GRPA treatment which clearly shows excitonic resonances. 

We find  excitonic modes in both the intra-valley and the inter-valley sectors. We find that the intra-valley excitons have a much larger binding energy. Upon increasing interaction strength, the excitons condense at zero momentum to give rise to a magnetic transition. The Weyl modes merely shift without opening up a gap. Our results show that Hubbard-like interactions are unlikely to give rise to the axionic CDW transition. This is consistent with results from cluster perturbation theory\cite{Krempa2014}.  

Excitonic modes have been experimentally seen in several graphene-like systems\cite{Wang2005,Ma2005,Dresselhaus2007,Shinar1999,Yang2007,Yadav2015}. 
Other two dimensional systems with a Dirac-like low energy description, transition metal dichalcogenides in particular, also host excitonic modes\cite{Singha2011,Park2017,Thygesen2017}. 

Our study shows that three dimensional Dirac systems are also highly conducive for excitonic modes. In particular, probes such as neutron scattering and photoabsorption could reveal excitonic resonances in candidate Weyl materials.

\acknowledgments 
It is a pleasure to thank G. Baskaran and S. A. Jafari for insightful discussions and comments. NSS thanks Subhankar Khatua and Prashanth Raman for discussions. We also thank Min-Fong Yang for pointing out errors in a previous version of the manuscript. 
\appendix
\section{Evaluating bare susceptibility}
\label{app.bare}
The sum in Eq.~\ref{eq.chibare} can be converted into an integral. In particular, the imaginary part of $\chi^{0}(\mathbf{q},\tilde{\omega})$ only receives contributions from points where the real part of the denominator vanishes. More precisely, $\chi^{0}(\mathbf{q},\tilde{\omega})$ counts the number of $\mathbf{k}$-points that satisfy $E_{\mathbf{k}+\mathbf{q}}+E_{\mathbf{k}}=\tilde{\omega}$. This leads to
\begin{eqnarray}
 &\operatorname{Im}(\chi^{0}(\mathbf{q},\tilde{\omega}))=\int_{\mid k\mid<k_{c}}d^{3}k \phantom{a}\delta[\tilde{\omega}-(E_{\mathbf{k}+\mathbf{q}}+E_{\mathbf{k}})].\phantom{ab}
\label{eq.ellipsoid}
\end{eqnarray}
To evaluate this, we consider a potential function $f_{\mathbf{q}}(\mathbf{k})=E_{\mathbf{k}+\mathbf{q}}+E_{\mathbf{k}}$. The delta function picks out an equipotential surface on which $f_\mathbf{q}$ takes the value $\tilde{\omega}$. This integral can be evaluated using methods that are typically used in density-of-states calculations, 
\begin{eqnarray}
 \operatorname{Im}(\chi^{0}(\mathbf{q},\tilde{\omega}))
 = \int_{\mathcal{E}} \frac{ds}{\vert \mathbf{\nabla}_{k}f_{\mathbf{q}}(\mathbf{k})\vert },
\label{eq.ellipsoid1}
\end{eqnarray}
where $\mathcal{E}$ denotes the equipotential surface in $\mathbf{k}$-space where $f_\mathbf{q}=\tilde{\omega}$, with $dS$ being its area element. The magnitude of the gradient in the denominator gives the density of states that are available in the vicinity of the point on the surface. As described in the main text, this surface is an ellipsoid. 

We evaluate this integral in spherical coordinates. Taking $\mathbf{q}$ to lie along the $z$-direction, we define polar and azimuthal angles $\theta$ and $\phi$. The condition $\{f_\mathbf{q}=\tilde{\omega}\}$ reduces to $\{k + \sqrt{k^2 + q^2 + 2kq\cos\theta} = \tilde{\omega}\}$, which determines $k$ as a function of $\theta$. We obtain  $k_\theta = \frac{\tilde{\omega}^2-q^2}{2(\tilde{\omega}+q\cos \theta)}$. The integral becomes
\begin{eqnarray}
\nonumber 
 \int_0^{2\pi} d\phi \int_{0}^\pi d\theta k_\theta sin \theta  \sqrt{k_\theta^2 d\theta^2 + dk_\theta^2}   \frac{\tilde{\omega}^2 + q^2 +2\tilde{\omega} q\cos\theta}{2(\tilde{\omega}+q\cos\theta)}  \\
= \frac{\pi(3\tilde{\omega}^2 - q^2)}{6}, q < \tilde{\omega} < (2k_c+q).\phantom{abc}
\end{eqnarray}
If $\tilde{\omega}$ were to be less than $q$ or greater than $2k_c +q$, then $ \operatorname{Im}(\chi_{\alpha\beta}^{0}(\mathbf{q},\tilde{\omega}))$ vanishes. The real part of the susceptibility can be evaluated using the Kramers-Kronig relation,
\begin{eqnarray}
\begin{aligned}
 &\operatorname{Re}(\chi^{0}(\mathbf{q},\tilde{\omega})) = \pi k_c^2+\\
& \frac{\pi}{6}\left[6(q+\tilde{\omega})k_c+(q^2-3\tilde{\omega}^2)\log\left\{\frac{q-\tilde{\omega}}{2k_c+q-\tilde{\omega}}\right\}\right].
 \end{aligned}
\end{eqnarray}
Close to the particle-hole continuum ($\tilde{\omega} \lesssim q$), the above expression can be approximated as
\begin{eqnarray}
 &\operatorname{Re}(\chi^{0}(\mathbf{q},\tilde{\omega})) =  \pi k_c^2+\frac{\pi}{6}\left[12 q k_{c}-2q^2\log\left\{\frac{q-\tilde{\omega}}{2k_c}\right\}\right].\phantom{ab}
 \end{eqnarray}

\section{GRPA expressions}
\label{app.GRPA}

To evaluate the bare susceptibility matrix, we first diagonalize the non-interacting Hamiltonian of Eq.~\ref{eq.H_TINI}. This is achieved by a unitary transformation, $\gamma_{\mathbf{k},v} \equiv U_{v,\sigma}(\mathbf{k})c_{\mathbf{k},\sigma}$. Here, $\gamma$'s are quasiparticle operators in the band basis and $U(\mathbf{k})$ is $2\times2$ unitary matrix. 
The diagonalized Hamiltonian is $Diag\{E_{\mathbf{k}},-E_{\mathbf{k}}\} $ where $E_{\mathbf{k}}=\sqrt{k_x^2+k_y^2+m(k_z)^2}$.

As described in the main text, we identify suitable bilinears for the inter-valley and intra-valley sectors separately. The expressions for the bare susceptibility matrix are given below.

\subsection{Inter-Valley}
\label{app.GRPAinter}
In the inter-valley sector, the bare susceptibility is given by the expression,
\begin{eqnarray}
\nonumber
\chi^0_{\mu\nu}(\mathbf{q},\omega)=\frac{1}{N}\sum_{\mathbf{k}\in L}\left[\frac{M^{\mu}(\mathbf{k},\mathbf{q})[M^{\nu}(\mathbf{k},\mathbf{q})]^{*}}{\omega+E(\mathbf{k+q})+E(\mathbf{k})}\right. \\
- \left. \frac{N^{\nu}(\mathbf{k},\mathbf{q})[N^{\mu}(\mathbf{k},\mathbf{q})]^{*}}{\omega-E(\mathbf{k+q})-E(\mathbf{k})}\right].
\label{eq.bare}
\end{eqnarray}
Here, the momentum $\mathbf{k}$ is summed over the `left' half of the Brillouin zone ($k_z<0$) to avoid double counting. The momentum $\mathbf{q}$ is restricted to the vicinity of $\mathbf{Q}$ so that we only consider inter-valley excitations. The indices $\mu$ and $\nu$ denote components of the vector of bilinears defined in the main text.
The non-zero elements of $\chi^0$ are obtained by plugging the following functions into Eq.~\ref{eq.bare}.
\begin{eqnarray}
\nonumber M^{\hat{\rho}/\hat{S_z}}&=&\frac{1}{2}\left[U_{12}^*(\mathbf{k}+\mathbf{q})U_{11}(\mathbf{k})\pm U_{22}^*(\mathbf{k}+\mathbf{q})U_{21}(\mathbf{k})\right]\\
\nonumber  &=&N^{\hat{\rho}^{\dagger}/\hat{S_z}^{\dagger}},\\
\nonumber M^{\hat{\rho}^{\dagger}/\hat{S_z}^{\dagger}}&=&\frac{1}{2}\left[U_{11}(\mathbf{k}+\mathbf{q})U_{12}^*(\mathbf{k})\pm U_{21}(\mathbf{k}+\mathbf{q})U_{22}^*(\mathbf{k}) \right]\\
\nonumber &=&N^{\hat{\rho}/\hat{S_z}},\\
\nonumber M^{\hat{S^{+}}}&=&U_{12}^*(\mathbf{k+q})U_{21}(\mathbf{k})=N^{(\hat{S^{+}})^{\dagger}},\\
\nonumber M^{\hat{S^{-}}}&=&U_{22}^*(\mathbf{k+q})U_{11}(\mathbf{k})=N^{(\hat{S^{-}})^{\dagger}},\\
\nonumber M^{(\hat{S^{+}})^{\dagger}}&=&U_{22}^*(\mathbf{k})U_{11}(\mathbf{k+q})=N^{\hat{S^{+}}},\\
M^{(\hat{S^{-}})^{\dagger}}&=&U_{12}^*(\mathbf{k})U_{21}(\mathbf{k+q})=N^{\hat{S^{-}}}.
\end{eqnarray}

\subsection{Intra-Valley}
\label{app.GRPAintra}
We use the 8-component vector of intra-valley bilinears as defined in the main text. The first four elements correspond to the left valley while the next four correspond to the right valley. The bare susceptibility matrix takes the form
\begin{eqnarray}
\chi^{0}_{\mu\nu}(\mathbf{q},\omega) = \left(  
\begin{array}{cc}
\chi^{L}_{\mu\nu}(\mathbf{q},\omega) & 0_{4\times 4} \\
0_{4\times 4} & \chi^{R}_{\mu\nu}(\mathbf{q},\omega)
\end{array}
\right).
\end{eqnarray}
It is block diagonal in the valley basis as perturbations within one valley cannot induce a response in the other. The valley-susceptibilites are given by,
\begin{eqnarray}
\nonumber
\chi^{L/R}_{\mu\nu}(\mathbf{q},\omega)=\frac{1}{N}\sum_{\mathbf{k}\in L/R}\left[\frac{M^{\mu}(\mathbf{k},\mathbf{q})[M^{\nu}(\mathbf{k},\mathbf{q})]^{*}}{\omega+E(\mathbf{k+q})+E(\mathbf{k})} \right. \\
\left. - \frac{N^{\nu}(\mathbf{k},\mathbf{q})[N^{\mu}(\mathbf{k},\mathbf{q})]^{*}}{\omega-E(\mathbf{k+q})-E(\mathbf{k})}\right].
\label{eq.bareL}
\end{eqnarray}
For each valley, $\mathbf{k}$ is summed is over the corresponding region ($k_z < 0$ or $k_z >0$). The momentum $\mathbf{q}$ is restricted to the vicinity of zero to ensure that we have intra-valley excitations.
The functions in Eq.~\ref{eq.bareL} are given by
\begin{eqnarray}
\nonumber M^{\hat{\rho}/\hat{S^{z}}}&=&\frac{1}{2}\left[U_{12}^*(\mathbf{k+q})U_{11}(\mathbf{k})\pm U_{22}^*(\mathbf{k+q})U_{21}(\mathbf{k})\right],\\
\nonumber M^{\hat{S^{+}}}&=&U_{12}^*(\mathbf{k+q})U_{21}(\mathbf{k}),\\
\nonumber M^{\hat{S^{-}}}&=&U_{22}^*(\mathbf{k+q})U_{11}(\mathbf{k}),\\
\nonumber N^{\hat{\rho}/\hat{S^{z}}}&=&\frac{1}{2}\left[U_{12}^*(\mathbf{k})U_{11}(\mathbf{k+q})\pm U_{22}^*(\mathbf{k})U_{21}(\mathbf{k+q}) \right],\\
\nonumber N^{\hat{S^{+}}}&=&U_{22}^*(\mathbf{k})U_{11}(\mathbf{k+q}),\\
N^{\hat{S^{-}}}&=&U_{12}^*(\mathbf{k})U_{21}(\mathbf{k+q}).
\end{eqnarray}
In each function, the momentum argument determines if it is evaluated in the left or right region.

\bibliographystyle{apsrev4-1} 
\bibliography{weyl}
\end{document}